\documentclass[
aps,%
10pt,%
final,%
notitlepage,%
oneside,%
twocolumn,%
nobibnotes,%
nofootinbib,%
superscriptaddress,%
noshowpacs,%
centertags]%
{revtex4} 
\usepackage{titlesec}
\titlespacing*{\subsection}{0pt}{8pt}{4pt}

\begin{document}

{\footnotesize ISSN 1063-7729

Astronomy Reports, 2012, Vol. 56, No. 12, pp. 915--930

DOI: 10.1134/S1063772912120013
} 

  \selectlanguage{english}

\title{Structure of CB 26 Protoplanetary Disk Derived from Millimeter Dust Continuum Maps}

\author{\firstname{V.}~\surname{Akimkin}}
\email{akimkin@inasan.ru}
\affiliation{Institute of Astronomy, Russian Academy of Sciences, Moscow, Russia}

\author{\firstname{Ya.}~\surname{Pavlyuchenkov}}
\affiliation{Institute of Astronomy, Russian Academy of Sciences, Moscow, Russia}

\author{\firstname{R.}~\surname{Launhardt}}
\affiliation{Max Planck Institute for Astronomy, Heidelberg, Germany}

\author{\firstname{T.}~\surname{Bourke}}
\affiliation{Harvard Smithsonian Center for Astrophysics, Cambridge, USA}


\begin{abstract}
Observations of the circumstellar disk in the Bok globule CB\,26 at 110, 230, and 270\,GHz
are presented together with the results of the simulations and estimates of the disk parameters. These
observations were obtained using the SMA, IRAM Plateau de Bure, and OVRO interferometers. The
maps have relatively high angular resolutions (0.4\,--\,1$^{\prime\prime}$), making it possible to study the spatial structure
of the gas--dust disk. The disk parameters are reconstructed via a quantitative comparison of observational
and theoretical intensity maps. The disk model used to construct the theoretical maps is based on the
assumption of hydrostatic and radiative equilibrium in the vertical direction, while the radial surface density
profile is described phenomenologically. The system of equations for the transfer of the infrared
and ultraviolet radiation is solved in the vertical direction, in order to compute the thermal structure of the
disk. The disk best-fit parameters are derived for each map and all the maps simultaneously,
using a conjugate gradient method. The degrees of degeneracy of the parameters describing the thermal
structure and density distribution of the disk are analyzed in detail. All three maps indicate the presence
of an inner dust-free region with a radius of approximately 35~AU, in agreement with the conclusions of
other studies. The inclination of the disk is 78$^\circ$, which is smaller than the value adopted in our earlier study
of rotating molecular outflows from CB 26. The model does not provide any evidence for the growth of dust
particles above $a_{\rm max}\approx0.02$\,cm.
\end{abstract}  
\maketitle 
  \section{Introduction}
The rapidly growing number of detected planetary
systems, which have very different orbital configurations
and orbit stars with masses varying over a wide
range\footnote{http://exoplanet.eu}, suggests that the formation of planets is a
widespread process in the Galaxy. Since gas--dust
disks are natural predecessors of planetary systems,
theoretical and observational studies of such disks
have become intense over the past decade.

According to current understanding, stars are
formed during the gravitational collapse of molecular
clouds \cite{2007ARA&A..45..565M}. Since clouds always have non-zero angular
momenta, the collapsing matter cannot directly
fall into a protostar, and fairly rapidly (over $\sim 10^4$~yr)
forms a circumstellar disk surrounded by an envelope~\cite{2005A&A...442..703H}. The angular momentum inside the disk is
redistributed so that the bulk of the matter falls onto
the protostar, and only a smaller portion moves outward,
carrying away angular momentum \cite{1974MNRAS.168..603L, 2011ARA&A..49..195A, 2004ARep...48..800T}. The characteristic evolutionary timescales of dust disks
around young single stars are several million years~\cite{2001ApJ...553L.153H}. At this time, a key mechanism is initiated that can
eventually lead to the formation of planets: the growth
of the size of the dust particles and their settling
toward the central plane of the disk. The enlargement
of dust particles with the subsequent formation of
planetesimals form the basis of the core accretion theory~\cite{1969QB981.S26}, which appears to describe the main
regime of planetary formation \cite{2012ApJ...745....4J}. The evolution of
dust in disks around young stars is confirmed by
their observed spectral energy distributions (SEDs)
in the IR and (sub)millimeter~\cite{2010ApJS..188...75M,2005ApJ...631.1134A}; together with the
high rate of occurence of exoplanets, this suggests a
protoplanetary nature for these disks. Additional important
factors affecting the evolution of young stars
are bipolar jets, outflows, and disk winds \cite{2007prpl.conf..277P}, which
play an important role in the final stages of evaporation
of the gaseous disk. The variety of controlling processes
and the wide range of their physical parameters
makes protoplanetary disks very interesting objects;
a complete understanding of their nature will be only possible through a synthesis of modern observations
and detailed simulations.

Observations of protoplanetary disks are difficult,
since these objects have relatively small sizes and low
temperatures, but a number of basic facts concerning
their evolution and structures have been more or less
reliably established \cite{2011ARA&A..49...67W}. Observations in the middle-IR can be used to determine the rate of occurence
and lifetime of these disks \cite{2009AIPC.1158....3M}. Their masses can
be determined using millimeter observations \cite{2010ApJ...725..430M}, and
the structures of protoplanetary disks can be reconstructed
using interferometric observations \cite{2011A&A...527A..27S}. A
number of tools have been developed as a theoretical
basis for understanding the physics of protoplanetary
disks, and for the interpretation of SEDs (see, e.g.,
the list of publicly available software codes in \cite{2008NewAR..52..145W}). However, the problem of degeneracy remains, when
observations are equally well described by several sets
of parameters, and a model can only provide limits
for these sets \cite{2007MNRAS.382.1707H}. Increasing the angular resolution
of observations can help to resolve this problem.
Spatially resolved observations are currently available
for more than one hundred and fifty protoplanetary
disks at wavelengths from the optical to the radio,
in both molecular lines and the continuum\footnote{http://circumstellardisks.org/}. These
high-quality observational data require adequate interpretations
and new approaches to deriving as much
information as possible about the physical and chemical
structure of the disk. The successful determination
of the parameters of a number of protoplanetary
disks using multi-frequency, spatially resolved
observations \cite{2011ApJ...741....3K,2012arXiv1205.4901M} and the operation of the ALMA
interferometer make this area promising.

The aim of our study is to determine a self-consistent
physical structure for the protoplanetary
disk located at the edge of the Bok globule CB~26.
This disk is about a hundred AU in size, has a
mass of $0.1M_{\odot}$, and belongs to Class I of young
stellar objects according to the classification of \cite{1987IAUS..115....1L}. The disk in CB~26 has been well studied and modeled~\cite{2001ApJ...562L.173L, 2004ApJ...617..418S}. A rotating outflow responsible for the
removal of angular momentum from the disk was
detected in \cite{2009A&A...494..147L}. A central region $45\pm5$~AU in size
where dust is depleted was later discovered in the disk using
spatially resolved observations at 230 GHz \cite{2009A&A...505.1167S}. The
best-fit model was derived by comparing observed
images and synthetic maps. However, due
to the computationally intensive method used for
the radiative-transfer calculations, a step-by-step
method was used to determine the parameters, and
possible degeneracy of the model parameters was
not analyzed. Another shortcoming of the above work is
the phenomenological law used for the disk density distribution in the vertical direction. Our aim is to develop
new tools to reconstruct disk structure and advanced routine to identify best-fit model parameters and their degeneracies.
We aim to determine to what degree the modeling
of mm maps of CB~26 are reproducible and reliable
in the light of (a) new observations with the IRAM
Plateau de Bure Interferometer (PdBI), (b) a refined
model for the vertical density distribution, and (c) a
more well-grounded parameter search technique. 

The paper contains four sections, describing the
observations, a physical model for the disk, the procedures
used to obtain the synthetic maps, and our
searches for the best-fit model. The main results and
conclusions are given at the end of the paper.

    \section{OBSERVATIONS AND DATA REDUCTION}
 \subsection{OVRO observations} \label{sec_obs_ovro}
 CB~26 was observed with the Owens Valley Radio
Observatory (OVRO) interferometer between January
2000 and December 2001. The 1.3 and 2.7~mm
continuum emission was observed simultaneously in
2-GHz bands using an analog correlator, except in
the highest-resolution configuration, where the 4-GHz capability of the new 1~mm receivers was used.
Four configurations of the six 10.4~m antennas provided
baselines in the range 10\,--\,170\,k$\lambda$\ at 2.7~mm
(110 GHz) and 10\,--\,370\,k$\lambda$\ at 1.3~mm (236 GHz). The
average system noise temperatures of the He-cooled
SIS receivers were 300\,--\,400 K at 110 GHz and 300\,--\,600 K at 236~GHz. The observing parameters are
summarized in Table~\ref{tab-obs}.

 \begin{table*}[ht]
 \centering 
 \caption{Millimeter interferometric observations of CB~26 in the continuum: $\lambda_0$ and $\nu_0$ are the central wavelength and
frequency, $\theta$ the width of the antenna main beam, $a$ and $b$ the semi-major and semi-minor axes of the interferometer
beam, PA the position angle of the beam major axis measured from the northern protoplanetary disk axis in the clockwise
direction, $\sigma_{\nu}$ the rms map uncertainty in mJy/$\Omega_{\nu}$, and $\Omega_{\nu} = 1.133ab$ the effective solid angle of the beam.}
      \label{tab-obs} 
 \begin{tabular}{lrcccccc}
 \hline \hline \noalign{\smallskip}
 Telescope                  &
                   &
 $\lambda_0$           &
$\nu_0$              &
 $\theta$               &
 $a\times b$     &
 PA &
$\sigma_{\nu}$             \\
                       &
                       &
                       &
                       &
                  &
  HPBW                  &
                   &
                       \\
                       &
                       &
 [mm]                  &
 [GHz]                 &
 [$^{\prime\prime}$]              &
 [$^{\prime\prime}$]              &
 [deg]                 &
 [mJy/$\Omega_{\nu}$]            \\
 \hline  
 OVRO & 2001     & 2.7  & 110  &  65 & $1.20\times 0.88$ & 93    & 0.3  \\
 PdBI & 2005/08  & 3.4  & 88   &  57 & $3.18\times 3.02$ & 102   & 0.1  \\
 OVRO & 2001     & 1.3  & 236  &  25 & $0.58\times 0.40$ & 92    & 0.6  \\
 PdBI & 2005/09  & 1.3  & 230  &  22 & $0.46\times 0.39$ & 22    & 0.3  \\
 SMA  & 2006     & 1.1  & 270  &  47 & $1.00\times 0.84$ & $-$89 & 2.7 \\
 \hline\noalign{\smallskip}
 OVRO+PdBI$^{a)}$ & 2001--08 & 2.7  & 110  &  60 & $1.37\times 1.05$ & 124   & 0.15  \\
 OVRO+PdBI$^{a)}$ & 2001--09 & 1.3  & 230  &  22 & $0.39\times 0.36$ & ~75   & 0.5   \\
 SMA$^{a)}$       & 2006    & 1.1  & 270  &  47 & $1.00\times 0.84$ & $-$57 & 2.7   \\
 \hline
 \end{tabular}
 \newline
 a) Composite maps have been rotated by 32$^\circ$ in order for the plane of the disk to align with the $x$ axis.
 \end{table*}

 The amplitude and phase
calibration were based on frequent observations of
a nearby quasar. The flux densities were calibrated
using observations of Uranus and Neptune, yielding
relative uncertainties of 20\%. The raw data were
calibrated and edited using the MMA software package \cite{1993PASP..105.1482S}. The mapping and data analysis were carried
out using the MIRIAD package \cite{1995ASPC...77..433S}. The final maps
were obtained by combining the OVRO and IRAM
PdBI data.

 \subsection{IRAM PdBI observations} \label{sec_obs_pdbi}
Observations of CB~26 were carried out with the
IRAM Plateau de Bure Interferometer in November
2005 (D configuration with five antennas) and
December 2005 (C configuration with six antennas;
project PD0D). Two receivers were used simultaneously,
tuned to single side bands at 89.2~GHz and
230.5~GHz, respectively. Higher resolution observations
at 230 GHz were carried out in January 2009
(B configuration with six antennas) and February
2009 (A configuration with six antennas; Project
S078). Further observations at 86.7~GHz were obtained in November 2008 (C configuration) and
March 2009 (D configuration; project SC1C) with
six antennas, using the new 3~mm receivers. Several
nearby phase calibrators were observed during
each track to determine the time-dependent complex
antenna gains. The correlator bandpass was calibrated
using 3C~454.3 and 3C~273, and the absolute
flux-density scale was derived from observations of
MWC~349. The flux calibration uncertainty is estimated
to be $\leq$20\% at both wavelengths. The observing
parameters are summarized in Table~\ref{tab-obs}.

The raw data were calibrated and imaged using the
latest version of the GILDAS\footnote{http://www.iram.fr/IRAMFR/GILDAS} software. The final
maps were obtained by combining the OVRO and
PdBI data.

 \subsection{SMA Observations} \label{sec_obs_sma}
Observations with the Submillimeter Array\footnote{The Submillimeter Array is a joint project between the
Smithsonian Astrophysical Observatory and the Academia
Sinica Institute of Astronomy and Astrophysics and is
funded by the Smithsonian Institution and the Academia
Sinica.} (SMA, \cite{2004ApJ...616L...1H}) were made on December 6, 2006 (extended
configuration) and December 31, 2006 (compact
configuration), covering frequencies from 267 to
277~GHz in the lower and upper sidebands, respectively,
and providing baselines of 12\,--\,62\,k$\lambda$. The
typical system temperatures were 350\,--\,500~K. The
quasar 3C~279 was used for the bandpass calibration,
and the quasars B0355+508 and 3C~111 for the gain
calibration. Uranus was used for the absolute flux
calibration, which is accurate to 20\%\,--\,30\%. The 1.1~mm continuum map was constructed using line-free
channels in both sidebands. The main observational
parameters are summarized in Table~\ref{tab-obs}. The raw
data were calibrated using the IDL MIR package~\cite{Qi2005MIR}
and visualized using a package MIRIAD.

  \section{PHYSICAL MODEL}
A global aim of this paper is to develop a set of
software for determining the parameters of protoplanetary
disks using observed millimeter maps. The
reconstruction of the disk parameters is based on
a quantitative comparison of synthetic and observed
images of the disk. Mathematically, the problem is reduced
to searching for the minimum of the parameter-dependent
function describing the difference between
the observed and synthetic maps. In this Section,
we present the protoplanetary disk model used to
construct the synthetic maps.

The main sources of disk heating in the model are
the radiation of the central star and viscous heating
(important only in dense central regions), while the
source of cooling is thermal radiation in the continuum.
Determining the disk's thermal structure is reduced to the problem of radiative transfer in a dusty
medium, since the dust contributes mostly to the
opacity of the matter in protoplanetary disks. The
density distribution in a low-mass disk around a single
star can be considered to be axially symmetric in
a first approximation (this is also valid for peripheral
regions around close binaries). The thickness of the
disk increases fairly rapidly with radius, so that the
central star directly illuminates not only the inner
edge of the disk, but also its peripheral parts \cite{1987ApJ...323..714K}. Moreover, the dust temperature is mainly determined
by the radiation incident from the disk surface, i.e., by
the vertical radiation transfer, since the optical depth
rapidly increases in the radial direction. This makes it
possible to decompose the two-dimensional problem
and construct a so-called 1+1D model
of the protoplanetary disk, in which the vertical and
radial structures of the disk are determined independently \cite{1998ApJ...500..411D,2002A&A...389..464D}.
 
The radial distribution of the surface density in our
model is given by the analytical expression
\begin{equation}
 \Sigma_{\rm T}=\Sigma_0^{\rm gas}\left(\frac{R}{1\,\mbox{AU}}\right)^p,
 \label{densp}
\end{equation}
whose parameters are determined from observational
data. The disk was considered to be hydrostatic in
the vertical direction, and the gas-density distribution
was calculated based on its temperature. The
dust and gas were assumed to be well mixed, so
that the ratio of their densities did not change over
the disk, and was equal to 0.01. It was additionally
assumed that the gas and dust temperatures were
equal; this equality is achieved in dense regions due
to effective collisions between the dust particles and
molecules. In the disk atmosphere, the gas can be
much more hotter than the dust \cite{2004A&A...428..511J}, but there is only
a small mass of dust in this region. Therefore, the
assumption that the dust and gas temperatures are
equal is justified when obtaining synthetic maps of the
disk in the mm, where the disk is optically thin and
the radiation-intensity distribution reflects the dust density
distribution.

The calculation of radiation transfer in a protoplanetary
disk is a computationally intensive task due
to the high optical depths for photons in the UV
(up to $\sim10^5$). Since the computation time required
to calculate one model is critical to searches for the
best-fit model, we developed a fast method for calculating
the dust temperature in the disk with acceptable
accuracy. This was achieved through an accurate
transition to a two-frequency approximation and the use
of the Schuster--Schwarzschild and Eddington approximations.
We tested the method and compared
it with another algorithm for calculating the radiative
transfer, in which the dust opacity coefficients are
functions of the wavelength. The transition from
monochromatic opacities to mean opacities makes it
possible to appreciably reduce the computational time
without losing accuracy when calculating the dust
temperature.

\subsection{Momentum Equations
for the IR Radiation Transfer} 

Let us consider a ring in the disk at the radius $R$.
The  gas surface density at this radius, in a layer between the equatorial plane and the surface of the
disk, is $\Sigma_{\rm T}$.  The key assumption of the model is
that the disk does not radiate in the UV. Thus, we
divided the spectral range into UV and IR parts and
describe them separately. It was also assumed that
the disk is geometrically thin and the plane-parallel
approximation is valid. If $\Sigma = \int\limits_{0}^{z} \rho(\tilde{z}) d\tilde{z}$ is the surface
density in a layer from 0 to $z$, then the momentum 
equations for IR radiation transfer in the grey and
Eddington approximations can be represented in the
form
\begin{align}
&\dfrac{dF}{d\Sigma} = c \kappa_P (B - E) \label{mom1},\\
&F = - \dfrac{c}{3\kappa_R} \dfrac{dE}{d\Sigma} \label{Edd},
\end{align}
where $F$ [erg cm$^{-2}$ s$^{-1}$] and $E$ [erg cm$^{-3}$] are the integrated
energy flux and energy density, $B = aT^4$ is
the integrated density of black-body radiation, and $\kappa_P$ and $\kappa_R$ [cm$^2$ g$^{-1}$]
are the Planck and Rosseland mean
opacities:
\begin{align}
\label{kappas1}
&\kappa_P = \dfrac{\int\limits_{0}^{\infty} \kappa_{\nu}^{\rm abs} B_{\nu}d\nu}
{\int\limits_{0}^{\infty} B_{\nu}d\nu}, \\\label{kappas2}
& \dfrac{1}{\kappa_R} = \dfrac{\int\limits_{0}^{\infty}\dfrac{1}{\left(\kappa_{\nu}^{\rm abs}+\kappa_{\nu}^{\rm sca}\right)}
\dfrac{\partial B_{\nu}}{\partial T}d\nu}{\int\limits_{0}^{\infty}\dfrac{\partial B_{\nu}}
{\partial T}d\nu}.
\end{align}
Here, $\kappa_{\nu}^{\rm abs}$ and $\kappa_{\nu}^{\rm sca}$ are the monochromatic coefficients
of true absorption and scattering. Note that
$\kappa_P$ and $\kappa_R$ are functions of the dust temperature.

\subsection{Balance Equation and the Heating Functions} 

The equations considered must be closed by an
energy-balance equation. The total radiative flux in
the IR is due to heating of the surrounding medium.
The change in this flux in the vertical direction is
described by the formula
\begin{equation}
\frac{dF}{d\Sigma} = S_{\rm star} + S_{\rm acc}.
\label{flux}
\end{equation}
Here, $S_{\rm star}$ and $S_{\rm acc}$ [erg s$^{-1}$ g$^{-1}$] represent the heating
due to stellar UV radiation and gas accretion. The
stellar radiation heating function is 
\begin{equation}
S_{\rm star} = 4\pi\kappa_{P}^{\rm uv} J_{\rm uv},
\end{equation}
where $\kappa_{P}^{\rm uv}$ is the Planck mean true absorption coefficient
in the UV per unit mass, and $J_{\rm uv}$ is the UV intensity averaged over angles and frequencies. We
used the two-flux Schuster--Schwarzschild approximation
to calculate
$J_{\rm uv}$, assuming that the disk does
not radiate in the UV, and only absorbs and scatters
the stellar radiation. The corresponding system of
equations has the form
\begin{align}
&\dfrac{dF_{\rm uv}}{d\Sigma} = - 4\pi \kappa_{P}^{\rm uv} J_{\rm uv}, \\
&F_{\rm uv}= - \dfrac{4\pi}{4\kappa_{F}^{\rm uv}} \dfrac{dJ_{\rm uv}}{d\Sigma},
\end{align}
where $\kappa_{F}^{\rm uv}$is the mean flux extinction coefficient (true
absorption and scattering), and $F_{\rm uv}$ is the UV flux.

The boundary conditions at the surface of the disk
can be written in the form
\begin{equation}
F_{\rm uv} = 2\pi J_{\rm uv} - 2\pi J_{\rm uv}^{-},
\end{equation}
where $J_{\rm uv}^{-}$ is the intensity of the stellar radiation averaged over a
hemisphere:
\begin{equation}
J_{\rm uv}^{-}=\dfrac{f}{\pi}\cdot L_{\rm star}/(4\pi R^2).
\end{equation}
 The coefficient $f$ determines the portion
of the stellar radiation intercepted by the disk. Generally, $f$ depends on the inclination of the disk surface
relative to the incident radiation. It is natural to
combine the input model parameters $f$ and $L_{\rm star}$ into
a single parameter $f\cdot L_{\rm star}$, representing the portion of
the stellar luminosity intercepted by the disk. 

The above system of equations can be solved analytically,
if the opacities do not depend on $\Sigma$, or
numerically, e.g., using the finite difference method.
The UV opacities are defined as follows:
\begin{align}\label{kappas3}
&\kappa_P^{\rm uv} = \dfrac{\int\limits_{0}^{\infty} \kappa_{\nu}^{\rm abs} B_{\nu}d\nu}
{\int\limits_{0}^{\infty} B_{\nu}d\nu}, \\\label{kappas4}
&\kappa_F^{\rm uv} = \dfrac{\int\limits_{0}^{\infty} \left( \kappa_{\nu}^{\rm abs} + \kappa_{\nu}^{\rm sca}\right) B_{\nu}d\nu}
{\int\limits_{0}^{\infty} B_{\nu}d\nu}. 
\end{align}
Here, the Planck function $B_{\nu}$ depends on the stellar
temperature. Assuming that the gravitational energy
release is proportional to the density, the heating
function due to gas accretion is 
\begin{equation}
S_{\rm acc} = \frac{G\dot{M}M_{\rm star}}{4 \pi \Sigma_{\rm T} R^3 },
\end{equation}
where $\dot{M}$ is the accretion rate, and $M_{\rm star}$ is the stellar
mass.

\subsection{Solution of the System of Equations
for the Thermal Structure}

The above heating functions depend only on the
surface density. Let us assume for the moment that $\kappa_P$ and $\kappa_R$ are only functions of the surface density
(they are functions of temperature as well, but this
problem will be solved further using an iteration process).
In this case, we can obtain an analytical solution
for $T(\Sigma)$.
The temperature can be determined
from Equations~\eqref{mom1} and~\eqref{flux}, i.e., from the relationship
\begin{equation}
c \kappa_P (aT^4-E) = S_{\rm star}+S_{\rm acc}.
\end{equation}
It is necessary to find $E(\Sigma)$ in order to obtain $T(\Sigma)$. We integrate Equation~\eqref{flux} and obtain $F = F(\Sigma)$. Then, $F(\Sigma)$ is substituted in Equation~\eqref{Edd}, and the latter is integrated as $\int\limits_{0}^{\Sigma}d\Sigma^{\prime}$. As a result, we obtain an equation with the
known function $\delta_E(\Sigma)$
\begin{equation}
E(\Sigma) - E(0) = \delta_E(\Sigma).
\label{Erad}
\end{equation}
We must know the radiation energy density $E(0)$ in
the equatorial plane of the disk in order to reconstruct $E(\Sigma)$ from this relationship. The radiation density
can be found from boundary conditions. If the IR
radiation from the disk is isotropic at the disk surface,
i.e., at  $\Sigma$=$\Sigma_{\rm T}$,
whereas the IR radiation from the star
can be neglected compared to the radiation from the
disk itself, then
$F(\Sigma_0) = \dfrac{1}{2} cE(\Sigma_{\rm T})$. Therefore, the
radiation energy density at the disk surface is
\begin{equation}
E(\Sigma_{\rm T}) = \dfrac{2F(\Sigma_{\rm T})}{c}.
\end{equation}
The formulas for $E(0)$ are obtained from~\eqref{Erad} for $\Sigma=\Sigma_{\rm T}$:
\begin{equation}
E(0) = E(\Sigma_{\rm T}) - \delta_E(\Sigma_{\rm T}).
\end{equation}

When constructing the solution, we assumed that $\kappa_P$ and $\kappa_R$ are functions only of the surface density,
not the temperature. An iterative process was implemented
to solve this problem. After the thermal structure
was found for the given opacities, we recalculated
the opacities using information on the temperature
profile $T(\Sigma)$. The new opacities were then used to
recalculate a refined thermal structure. In practice,
this iterative process converged over 8\,--\,10 steps.

\subsection{Equation of Hydrostatic Equilibrium}

The equation of hydrostatic equilibrium for a geometrically
thin disk has the form
\begin{equation}
\dfrac{1}{\rho}\dfrac{d}{dz} (c_T^2 \rho) = - \dfrac{GM_{\rm star}}{R^3}z, 
\end{equation}
where $c_T^2$is the sound speed. Since $T$, and therefore $c_T^2$ as a function of $\Sigma$, (but not as function of $z$) is
known, the latter equation should be written in the
form
\begin{align}
&\dfrac{dz}{d\Sigma} = \dfrac{1}{\rho} \label{tot1},\\
&\dfrac{d}{d\Sigma} (c_T^2 \rho) = - \dfrac{GM_{\rm star}}{R^3}z. \label{tot2}
\end{align}
This system of equations can be solved using the
explicit integration scheme if the density $\rho(0)$ in the
equatorial plane is known; this should correspond to
the boundary condition
$\rho(\Sigma_{\rm T}) = \rho_{\rm ext}$, where $\rho_{\rm ext}$ is the
given density at the disk surface. To search for the
required value of $\rho(0)$, we used a combination of the
shooting method and bisection method. The solution
$z=z(\Sigma)$ and $\rho=\rho(\Sigma)$ of the system of Equations~\eqref{tot1}--\eqref{tot2} can be used to obtain the required
relationships $\rho=\rho(z)$ and $T=T(z)$.

The choice of a grid of spatial coordinates is an
important part of the model. A logarithmic grid is
a natural choice for the radial direction. This grid
smoothly traces the density gradients in the case of a
power-law parametrization of the surface density~\eqref{densp}.
The left and right boundaries of the radial grid are
the inner $R_{\rm in}$ and outer $R_{\rm out}$ radii of the disk.The
choice of a grid in the vertical direction is a more
complex task. If we knew the density distribution
in the vertical direction {\it a~priori}, we could place the
grid nodes so that the density would change by no
more than some specified factor between adjacent
cells. However, the density profile is not known {\it a~priori}. On the other hand, the density distribution can
be found analytically for a vertically isothermal disk
with a specified temperature. Therefore, we used the
grid for an isothermal disk to calculate the density
distribution in a nonisothermal disk with a similar
temperature.

In practice, calculation of disk model on a grid of 500$\times$500 cells takes two to three minutes using a 3~GHz processor. Figure~\ref{Fig_PhysStructure} shows examples of the
density and temperature distributions for one model
of the disk in CB~26.
  \section{CONSTRUCTION OF SYNTHETIC MAPS}

After model temperature and density distributions
in the disk have been constructed, it is necessary
to calculate the spatial distribution of the radiation
intensity $I_{\nu}$ coming to the Earth for a given disk
orientation in the plane of the sky, i.e., to construct
a theoretical image. The theoretical maps were constructed
using the tracing routine from the NATALY
software~\cite{2011ARep...55....1P}. The scattering was neglected, since the two-frequency disk model does not enable calculation
of the distribution of the spectral radiation
intensity $J_{\nu}$. Note, however, that scattering is negligible
compared to thermal radiation in the millimeter
wavelengths.

We must know the monochromatic opacities $\kappa_{\nu}^{\rm abs}$ and $ \kappa_{\nu}^{\rm sca}$ to construct theoretical maps and obtain
the mean opacities in \eqref{kappas1}, \eqref{kappas2},
\eqref{kappas3}, and \eqref{kappas4}. 
The
monochromatic opacities were calculated based on
the true absorption efficiency factor $Q_{\rm abs}(\nu,a)$ and
the scattering efficiency factor
$Q_{\rm sca}(\nu,a)$, calculated
using Mie theory:
\begin{eqnarray}
 \kappa_{\nu}^{\rm abs}=\frac{1}{\rho_{\rm d}}\int\limits_{a_{\min}}^{a_{\max}} Q_{\rm abs}(\nu,a) \pi a^2 f(a)\,da, \\
\kappa_{\nu}^{\rm sca}=\frac{1}{\rho_{\rm d}}\int\limits_{a_{\min}}^{a_{\max}} Q_{\rm sca}(\nu,a) \pi a^2 f(a)\,da.  
\end{eqnarray}
The dust-particle size distribution function $f(a)$ was
assumed to be a power-law, parametrized in terms of
the maximum and minimum particle sizes $a_{\rm max}$ and $a_{\rm min}$
and the slope $a_{\rm pow}$. An additional parameter
is the mass fraction of silicate $f_{\rm Si}$ in the mixture of
silicate and graphite dust particles.

To obtain a model disk image $I_{\nu}^{*}$ (synthetic map),
the ideal disk image $I_{\nu}$ must be convolved with the
beam:
\begin{equation}
    I_{\nu}^{*}(x,y)=\int\int I_{\nu}(x',y') W(x'-x,y'-y) \,{\rm d}x'  {\rm d}y'.
    \label{Eq_convolution}
\end{equation}
The full widths at half maximum of the beam $W(a,b)$ were determined using the major and minor axes ${\rm HPBW}_a$ and ${\rm HPBW}_b$,
and the position angle ${\rm PA}$:

\begin{eqnarray}
 &W(a,b)=\frac{1}{\pi H_x H_y}\nonumber\\
 &\times\exp{\left[-\left(\frac{a\cos\alpha-b\sin\alpha}{H_x}\right)^2\right]}\nonumber\\
 &\times\exp{\left[-\left(\frac{a\sin\alpha+b\cos\alpha}{H_y}\right)^2\right]},
    \label{Eq_RadPattern}
\end{eqnarray}
where
\begin{equation}
 H_x =\frac{{\rm HPBW}_a}{2\sqrt{\ln 2}}, \quad H_y =\frac{{\rm HPBW}_b}{2\sqrt{\ln 2}}, \quad \alpha=\frac{\pi}{2}+{\rm PA}.
\end{equation}

We focused on the development of a rapid and
accurate calculation technique, since the convolution
with the beam is the most resource-intensive step of the numerical integration. The multiple integral~\eqref{Eq_convolution}
reduces to an iterated integral that was
calculated using a composite quadrature formula with
a highest degree of accuracy (the composite Gaussian
quadrature); i.e., the entire integration interval
was divided into unequal subintervals, within each of
which a simple Gaussian quadrature was used. The
subintervals are searched for using a local double recalculation
method: a subinterval is divided into
two new subintervals only if it yields a significant
improvement in the integration accuracy. This adaptive
control of the integration accuracy enables us to
appreciably reduce the time required to compute~\eqref{Eq_convolution}.
  \section{SEARCH FOR THE BEST-FIT MODEL}
We used an analog of the reduced $\chi^2_{\rm red}$ as a criterion for agreement between
the calculated and observed images:
\begin{equation}
 \chi^2_{\rm red}=\frac{1}{N}\sum_{(x,y)}\frac{\left[I_{\nu}^{\rm obs}(x,y)-I_{\nu}^{\rm *}(x,y)\right]^2}{\sigma_{\nu}^2}, 
\label{Eq_Chi2}
\end{equation}
where $N$ is the number of degrees of freedom of the
model, which is equal to the difference between the
number of observational points and the number of
model parameters, if the model is a linear function
of these parameters. If the model is nonlinear, it is
difficult (or even impossible) to determine the number
of degrees of freedom, since the computation of one
model becomes fairly resource-intensive~\cite{2010arXiv1012.3754A}. We
took $N$ to be equal to the difference between the number
of map pixels (60$\times$60=3600 pixels for a 3$^{\prime\prime}$$\times$3$^{\prime\prime}$ map) and the number of free parameters~($<10$).

The reconstruction of the disk properties is reduced
to searching for the minimum $\chi^2_{\rm red}$
as a function
of the free model parameters. It is quite natural to
single out four groups of parameters describing the
following disk properties:
\begin{description}
 \item[I.] Density distribution:  
\begin{itemize}
 \item $R_{\rm in}$~--- radius of the inner dust-depleted region; 
 \item $R_{\rm out}$~--- outer radius of the disk;
 \item $\Sigma_0^{\rm gas}$~--- gas surface density at a radius of 1~AU;
 \item $ p$~--- index of the power-law function describing
the surface density.
\end{itemize}

 \item[II.] Thermal structure:
\begin{itemize}
\item $f L_{\rm star}$~--- part of star radiative energy intercepted
by the disk;
\item $M_{\rm star}\dot{M}$~--- product of the stellar mass $M_{\rm star}$ and
the accretion rate $\dot{M}$, describing the accretion heating.
\end{itemize}
 \item[III.] Disk position: 
\begin{itemize}
\item $i$~--- inclination between the plane of the disk and
the line of sight ($i=90^{\circ}$ corresponds to an edge-on
disk);
\item ${\rm PA}$~--- position angle of the major axis of the disk
image relative to the $x$ axis of the map;
\item $D$~--- distance to the disk;
\item $x_s$~--- shift of the central star position relative to
the map point (0,0) along the $x$ axis;
\item $y_s$~--- shift of the central star position relative to
the map point (0,0) along the $y$ axis.
\end{itemize}

\item[IV.] Dust particle properties: 
\begin{itemize}
\item $f_{\rm Si}$~--- mass fraction of silicate grains mixtured
with graphite grains;
\item $a_{\rm min}$~--- minimum size of the dust grains;
\item $a_{\rm max}$~--- maximum size of the dust grains;
\item $a_{\rm pow}$~--- index of the power-law dust grain size
distribution.
\end{itemize}
\end{description}

Some of these parameters can be kept fixed. Tests
have shown that synthetic maps depend only weakly
on the accretion rate. This is due to the fact that
viscous heating dominates over heating by the stellar
radiation in the inner, dense regions that are heated
more, and whose maximum radiation occurs in the
middle IR. The outer rarefied and cool regions of
the disk are primarily responsible for the radiation in
the millimeter. Therefore, it is not surprising that
synthetic millimeter maps are not sensitive to $\dot{M}$. Further, we can independently determine the position
angle PA of the major axis of the disk image in the
plane of the sky and the shifts of the image centers $x_s, y_s$ relative to the central star. We fixed the distance
to the disk to be  $D=140$\,pc, since there is reason
to believe that the disk in CB\,26 is a member of
the Taurus--Auriga star-forming region~\cite{2001ApJ...562L.173L}. It is
reasonable to take the same parameters for the silicate
and graphite particle-size distributions, since
it has been shown that their evolutions do not differ
appreciably~[J\"urgen Blum,
priv. comm.]. We are going to determine to what
degree the dust particles can grow, i.e., to determine
their maximum size $a_{\rm max}$; thus, for simplicity, we fixed
the minimum size to be $a_{\rm min}=5\cdot 10^{-7}$\,cm and the
distribution slope to be $a_{\rm pow}=-3.5$, corresponding
to the interstellar medium \cite{1977ApJ...217..425M}. 

After the above parameters are excluded from consideration,
the following set of eight free parameters
remains: $(R_{\rm in}, R_{\rm out},\Sigma_0^{\rm gas},p;f L_{\rm star};i;f_{\rm Si},a_{\rm max})$.
If these parameters were determined using
only spatially unresolved observations, this would
yield certain difficulties due to the small number of
degrees of freedom in the model. The coverage of the
frequency range considered would be determined by
only about 20 points, corresponding to the available
observations (HST, IRAS, Spitzer, Herschel, SMA,
SCUBA); therefore, attempts to fit an SED alone often results in significant degeneracy of the model
parameters. However, interferometric data provide
additional information about the object's structure,
making it possible to reduce the degeneracy, or, for
a number of parameters, even remove this problem
altogether.

\begin{figure*}{!Ht}
\setcaptionmargin{5mm}
\onelinecaptionsfalse
\includegraphics[angle=270,width=0.6\textwidth]{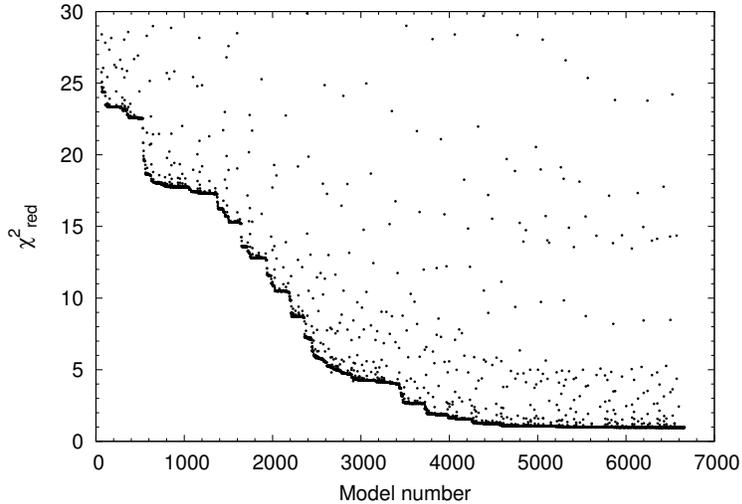}
\captionstyle{normal}
\caption{An example of searching for the best-fit model using the Powell method.}
\label{Fig_History}
\end{figure*}
We used the conjugate-gradient method of Powell \cite{Powell} to search for the minimum \eqref{Eq_Chi2}.This is a
fairly rapid algorithm that can solve the minimization
problem efficiently if the function considered has
''broad valleys''. These regions are characterized by
shallow, extended minima, such as are expected in the
problem considered here. For example, the radiation
flux from the disk is proportional to the disk mass and
the Planck function of the dust temperature, if the
disk is optically thin at a given wavelength. The same
flux could come from both a cool, massive disk and
a hotter, less massive disk. Therefore, the problem is
degenerate for the disk mass and mean dust temperature
in the disk, if the observations are not spatially
resolved. If interferometric observations are analyzed,
this problem can result in a mutual dependence between
the parameters for groups I and II. One of
our goals was to study the degree of degeneracy of
these parameters for spatially resolved observations
at a level of 1$^{\prime\prime}$ or better. The conjugate-gradient
method is optimal for solving such problems. Figure~\ref{Fig_History} presents an example of the convergence history of
this method for one observed map. The computation
time for one model is 5--20~min using one 3~GHz
processor.

We determined the parameter uncertainties as follows.
The boundaries of a confidence interval for a given parameter were taken to be the points at
which the value of $\chi^2_{\rm red}$ increases by unity compared
to the minimum value, if all the other parameters are
fixed and correspond to the minimum~\cite{1992nrfa.book.....P}. Thus,
if $\min \chi^2_{\rm red}=\chi^2_{\rm red} \left(a_1^{(0)},a_2^{(0)},a_3^{(0)},...\right)$, then the confidence interval for the parameter $a_1$ is the interval $\left(a_1^{(0)}-\sigma^-, a_1^{(0)}+\sigma^+\right)$, such that
 \begin{equation} \label{EqConfidIntrvl}
  \chi^2_{\rm red} \left(a_1^{(0)}\pm\sigma^{\pm},a_2^{(0)},a_3^{(0)},...\right)=\min \chi^2_{\rm red}+1.
 \end{equation}
The non-linearity of the model (more exactly, the
complexity of determining the number of degrees of
freedom~$N$) means that these confidence interval cannot
be taken to be accurate $1\sigma$ uncertainties, but we
consider them to be a measure of the parameter uncertainty,
close to the frequently used level of 68.3\%.

  \section{RESULTS}
\subsection{Best-Fit Models for Individual Maps}
\begin{figure*}[!ht]
\setcaptionmargin{2mm}
\includegraphics[angle=0,width=0.8\textwidth]{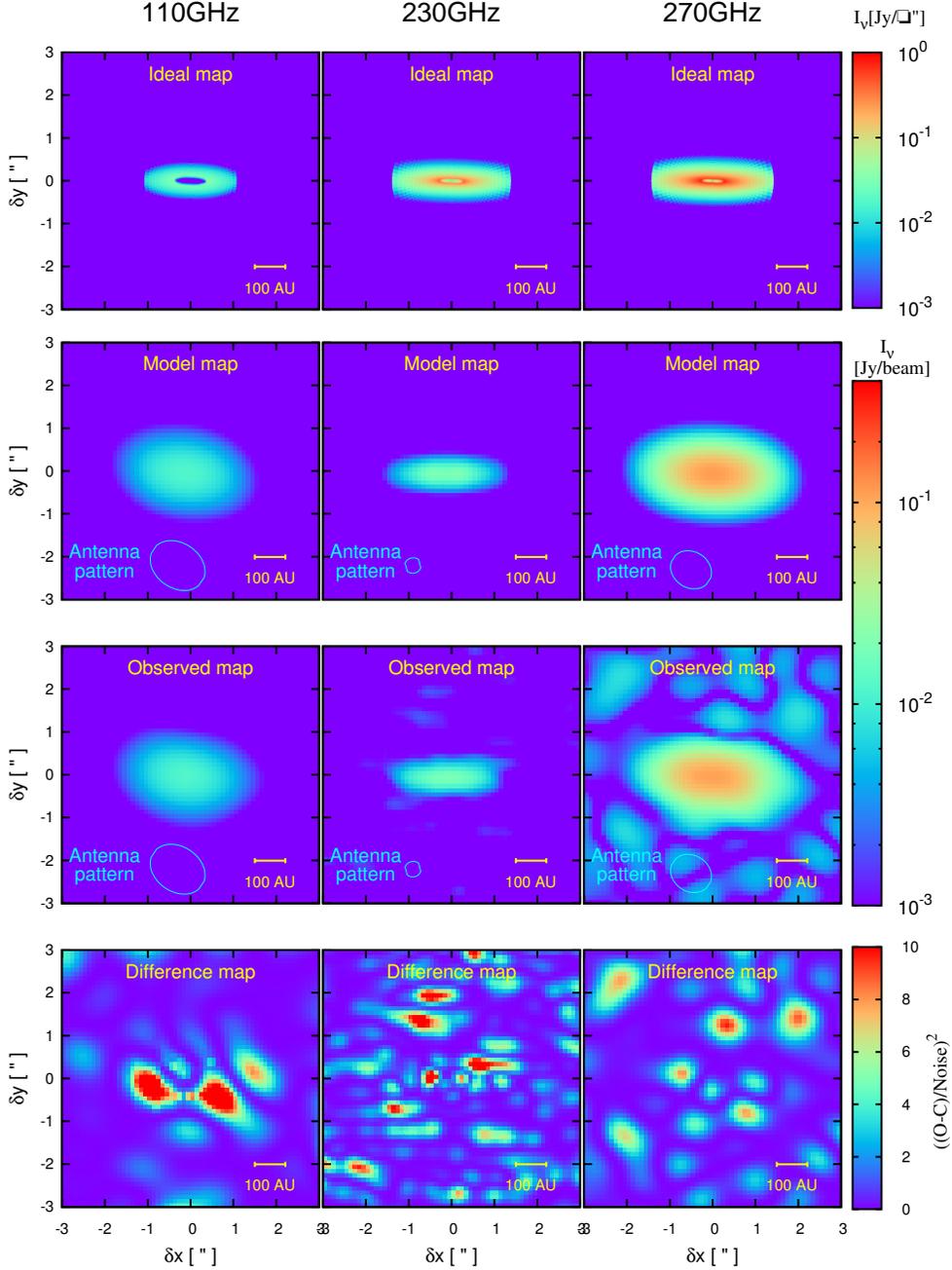}
\captionstyle{normal}
\caption{Maps of the disk radiation intensity at 110 GHz (left column), 230 GHz (central column), and 270 GHz (right column).
The upper row shows the ideal disk images for observations using a telescope with a point-like beam, the second row is the
ideal disk images convolved with the beam, the third row is the observed images, and the fourth row is the maps of the difference
between the observed (O) and model (C) images in noise units $\sigma_{\nu}$: $\rm ((O-C)/\sigma_{\nu})^2$.}
 \label{Fig_MapOfTheMaps}
\end{figure*}

In the first step of our study, we searched for best-fit
models for each of the three maps independently.
Figure~\ref{Fig_MapOfTheMaps} presents the results of this search, where the
disk images are given at 110 GHz (left), 230~GHz
(center), and 270 GHz (right). The upper row shows
ideal images that would be observed with a telescope
having a point-like beam. A prominent feature is
the presence of the central region devoid of dust,
with radii of approximately 55 AU at 110 GHz and
35 AU at 230 and 270 GHz. The maximum radiation intensity is reached at the edge of this region: 35, 280,
and 560 mJy/arcsec$^2$ at 110, 230, and 270 GHz,
respectively. The synthetic maps obtained by convolving
the ideal images with a beam are presented
in the second row. A comparison between the ideal
and convolved maps illustrates the difficulties arising in attempts to establish the disk sizes (morphology)
and orientations using the corresponding observed
maps. The maximum intensities in the synthetic
maps are 10, 20, and 110~mJy/$\Omega_{\nu}$ at 110, 230, and
270~GHz, respectively. The intensities are expressed
in terms of the effective beam solid angles $\Omega_{\nu}$, for convenience in comparing with the observational data
in the third row in Figure~\ref{Fig_MapOfTheMaps} ($\Omega_{110}=1.63$~arcsec$^2$, $\Omega_{230}=0.16$~arcsec$^2$, $\Omega_{270}=0.95$~arcsec$^2$).
The fourth row
in Figure~\ref{Fig_MapOfTheMaps} presents the distribution of $\chi^2_{\rm red}$ \eqref{Eq_Chi2} over
the image. On the whole, the differences in the
230 and 270~GHz maps are distributed randomly,
while the disk image seems to display structure at
110~GHz. This may provide evidence for the existence
of systematic difference between the theoretical and
observed maps at this frequency.

Numerical values of some model parameters corresponding
to the presented synthetic maps are given
in Table~\ref{Tab_BestFits}, together with the confidence intervals
calculated according to~\eqref{EqConfidIntrvl} and the minimum values $\chi^2_{\rm red}$. Note the high level of agreement between the
observed and synthetic maps: the mean deviation over
a map is less than 1.5 times the noise.
\begin{table*}
\setcaptionmargin{0mm} 
\caption{Parameters of the best-fit model.}
\label{Tab_BestFits}
\bigskip
    \begin{tabular}{l|c|c|c|c|c}\cline{1-6}
     Parameter             & 110 GHz                   & 230 GHz                  & 270 GHz                 & 230+270 GHz            & 110+230+270 GHz            \\\cline{1-6}
    $\chi^2_{\rm red}$     & 0.9                       & 1.3                      & 1.1                      & 1.7                     & 2.2                    \\
    $R_{\rm in}$, AU     & $54^{+7}_{-3}$            & $38^{+18}_{-12}$         & $36^{+15}_{-13}$         & 37$^{+16}_{-14}$        & 33$^{+13}_{-15}$       \\
    $R_{\rm out}$, AU    & $157^{+6}_{-12}$          & $204^{+161}_{-64}$       & $210^{+61}_{-41}$        & 222$^{+75}_{-57}$       & 207$^{+25}_{-32}$      \\
    $\Sigma_0^{\rm gas}$, g/cm$^2$   & $782^{+37}_{-78}$         & $809^{+206}_{-218}$      & $757^{+142}_{-126}$      & 710$^{+152}_{-147}$         & 865$^{+116}_{-114}$        \\
    $p$                    & $-0.64^{+0.02}_{-0.02}$   & $-0.96^{+0.06}_{-0.06}$  & $-0.87^{+0.04}_{-0.04}$  & $-0.81^{+0.04}_{-0.05}$       & $-0.69^{+0.03}_{-0.03}$      \\
    $i$, deg                    & $72^{+12}_{-13}$          & $78^{+12}_{-13}$         & $78^{+9}_{-21}$          & 78$^{+8}_{-16}$         & 82$^{+3}_{-17}$        \\
    $f\cdot L_{\star}$     & $0.03^{+0.004}_{-0.008}$  & $0.15^{+0.14}_{-0.09}$   & $0.12^{+0.06}_{-0.04}$   & 0.08$^{+0.05}_{-0.03}$      & 0.015$^{+0.005}_{-0.004}$     \\
    $f_{\rm Si}$           & $0.38^{+0.04}_{-0.10}$    & $0.49^{+0.28}_{-0.26}$   & $0.47^{+0.18}_{-0.20}$   &  0.59$^{+0.19}_{-0.20}$ &  0.37$^{+0.13}_{-0.13}$ \\
    $a_{\rm max}$, cm          & $<1.7\cdot 10^{-2}$       & $<1.4\cdot 10^{-2}$      & $<1.0\cdot 10^{-2}$      & $<1.3\cdot 10^{-2}$     & $<1.7\cdot 10^{-2}$     \\
    \end{tabular}
\end{table*}

\begin{figure}
\setcaptionmargin{5mm}
\includegraphics[angle=270,width=0.5\textwidth]{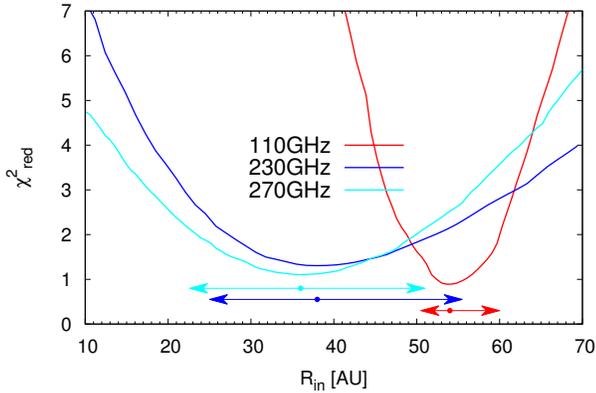}
\caption{Fit criterion $\chi^2_{\rm red}$ as a function of the inner disk radius $R_{\rm in}$. The other parameters were fixed at the values reached at the minimum $\chi^2_0=\min \chi^2_{\rm red}$. The arrows indicate the confidence intervals corresponding to $\chi^2_{\rm red}=\chi^2_0+1$, and the points on the
arrows correspond to the minimum.}
\label{Fig_Chi2R}
\end{figure}

We have analyzed in detail $\chi^2_{\rm red}$ as a function of
a number of important parameters: $R_{\rm in}$, $i$ and $a_{\rm max}$. Figure~\ref{Fig_Chi2R} shows the dependence
$\chi^2_{\rm red} (R_{\rm in})$ in the vicinity
of the minimum, keeping the remaining parameters
fixed. In this case also, the parameters obtained
for 110 GHz differ appreciably from those obtained for
the other two frequencies. The inner disk radius is approximately $ 37 \pm15$~AU at 230 and 270~GHz,
and approximately $55\pm5$~AU at 110~GHz. The presence
of a dust-free region with a size of $45\pm5$~AU was
suggested in~\cite{2009A&A...505.1167S} based on the presence of a central
plateau in the 230 GHz map. Figure~\ref{Fig_Chi2R} demonstrates
that neither the new observations obtained using the
PdBI (110 and 230 GHz) nor those used in~\cite{2009A&A...505.1167S} (obtained
using the SMA at 270 GHz) can be explained
without assuming the existence of a large, free-dust
central region. This region could have arisen due to
both dynamical effects, such as the sweeping up of
dust by a binary star, and evolutionary effects, such as
the formation of planetesimals in the central regions.
In this case, the absence of radiation from the central
region can be explained by inefficient reprocessing
of the radiation from the central star, rather than an
absence of matter. Such holes are often observed in
other disks~\cite{2008A&A...490L..15D, 2009ApJ...704..496B}.

\begin{figure}
\setcaptionmargin{5mm}
\onelinecaptionsfalse
\includegraphics[angle=270,width=0.5\textwidth]{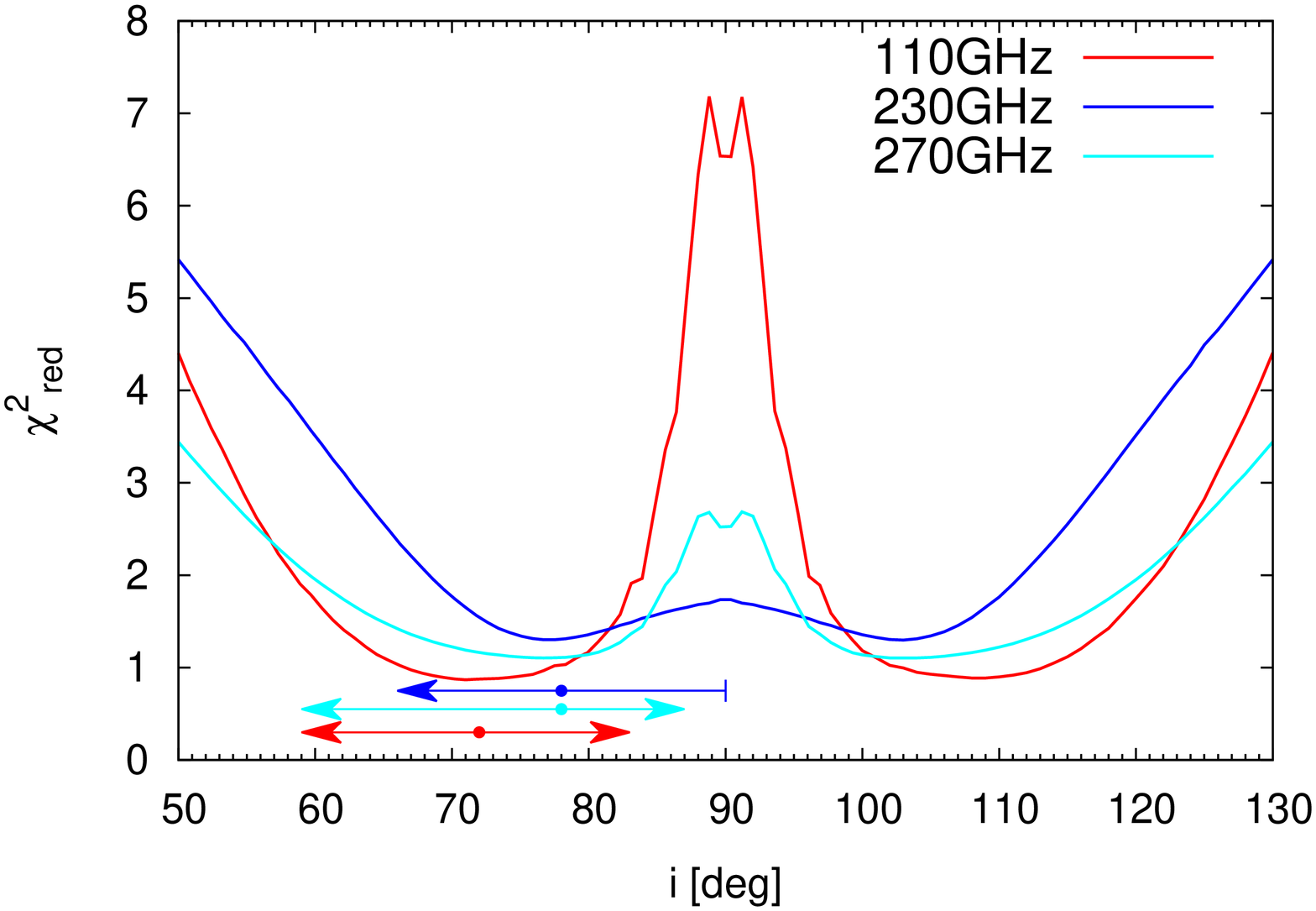}
\captionstyle{normal}
\caption{Fit criterion $\chi^2_{\rm red}$ as a function of the inclination angle $i$ of the disk to the plane of the sky. The other parameters were
fixed at the values reached at the minimum $\chi^2_0=\min \chi^2_{\rm red}$. The arrows indicate the confidence intervals corresponding to $\chi^2_{\rm red}=\chi^2_0+1$ and $i<90^{\circ}$ (this orientation follows from observations in molecular lines). The points on the arrows correspond
to the minimum. For $i<90^{\circ}$, the south pole of the disk axis is oriented toward the Earth.}
\label{Fig_Chi2i}
\end{figure}

The disk inclination $i$ is an important parameter
for modeling the bipolar outflow from CB~26. The
accuracy in this parameter can be estimated using Figure~\ref{Fig_Chi2i}, which presents the dependence $\chi^2_{\rm red} (i)$ for the
three frequencies. We show confidence intervals for $i<90^\circ$, since the southern edge of the disk is nearer
to the Earth, as is suggested by observations of the
outflow in molecular lines. The mean inclination over
the three maps is $76^\circ$. None of these millimeter maps
are sensitive to the edge of the disk that is nearest the
observer (the symmetry of $\chi^2_{\rm red} (i)$ relative to $90^\circ$). 

\begin{figure}
 \setcaptionmargin{5mm}
 \onelinecaptionsfalse
 \includegraphics[angle=270,width=0.5\textwidth]{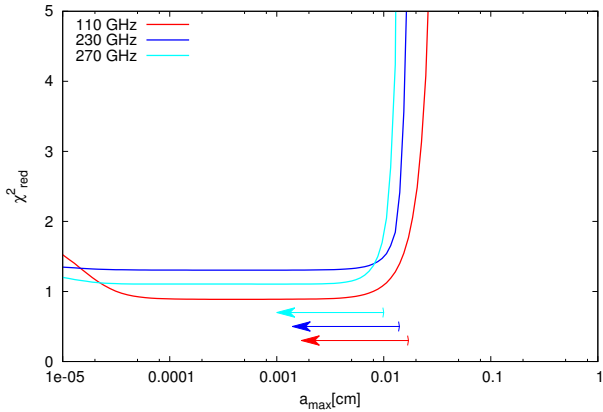}
 \captionstyle{normal}
 \caption{Fit criterion $\chi^2_{\rm red}$ as a function of the maximum size of the dust particles  $a_{\rm max}$. The other parameters were fixed at the
values reached at the minimum $\chi^2_0=\min \chi^2_{\rm red}$. The arrows indicate the confidence intervals corresponding to $\chi^2_{\rm red}=\chi^2_0+1$.}
 \label{Fig_Chi2amax}
 \end{figure}

Figure~\ref{Fig_Chi2amax} shows the dependence
$\chi^2_{\rm red} (a_{\rm max})$. The
synthetic maps are not sensitive to the maximum size
of the dust particles in the range from interstellar
values to $10^{-2}$~cm. Our model suggests an upper limit
to the maximum size of the dust particles in CB~26 of $0.02$~cm.
\begin{figure*}[!Ht]
\onelinecaptionsfalse
\includegraphics[angle=270,width=1.0\textwidth]{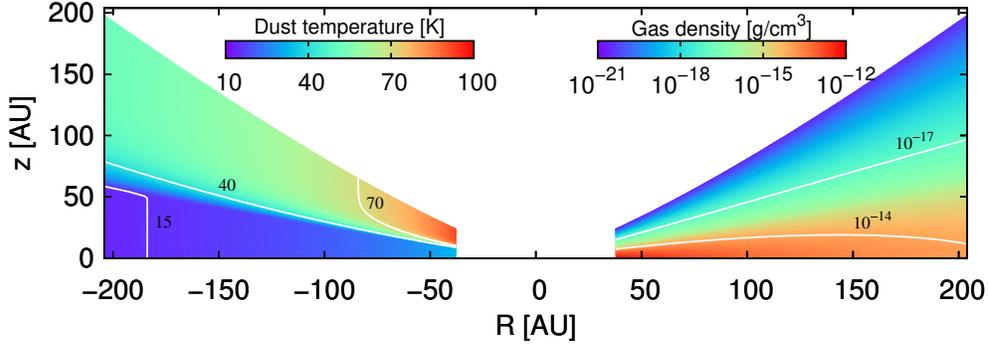}
\captionstyle{normal}
\caption{Dust temperature (left) and gas density (right) distributions for the disk models whose best-fit parameters correspond
to the observed maps at 230 and 270 GHz.}
\label{Fig_PhysStructure}
\end{figure*}

\subsection{Combined Model}

The fifth column in Table~\ref{Tab_BestFits} presents the model
parameters determined using the 230 and 270 GHz
maps jointly. The maps at these two frequencies
obtained with different angular resolutions using different
instruments yield quite similar disk parameters,
whereas the 110 GHz map gives different parameters
and shows the presence of correlated features in
the residuals (Figure~\ref{Fig_MapOfTheMaps}). Therefore, we excluded the
110 GHz map from our final search for the disk parameters.
The mean value of $\chi^2_{\rm red}=(\chi^2_{\rm 230}+\chi^2_{\rm 270})/2$
was used for $\chi^2_{\rm red}$.
The results obtained using all
three maps are given in the last column of Table~\ref{Tab_BestFits} for
comparison.

The physical structure of the disk corresponding
to the joint minimum
$\chi^2_{\rm red}$ is shown in Figure~\ref{Fig_PhysStructure}. The
dust temperature varies from 10~K in the peripheral
regions near the equatorial plane to 100~K in the atmosphere
of the disk at its center. The maximum gas
density in the equatorial plane reaches $10^{-12}$~g/cm$^3$. The normalized surface density, power-law index, and portion of the stellar radiation intercepted by the disk
are $\Sigma_0^{\rm gas}=710$~g/cm$^2$,
$p=-0.8$, and $f L_{\star}=0.08 L_{\odot}$. The full optical depth of the disk is close to unity at
all three frequencies. The maximum optical depth
is reached in the vicinity of the disk inner boundary,
and is equal to 0.35, 0.6, and 0.9 for 110, 230, and
270 GHz, correspondingly.

\subsection{Model Degeneracy}

\begin{figure}[!Ht]
  \setcaptionmargin{5mm}
  \onelinecaptionsfalse
  \includegraphics[angle=0,width=0.7\textwidth]{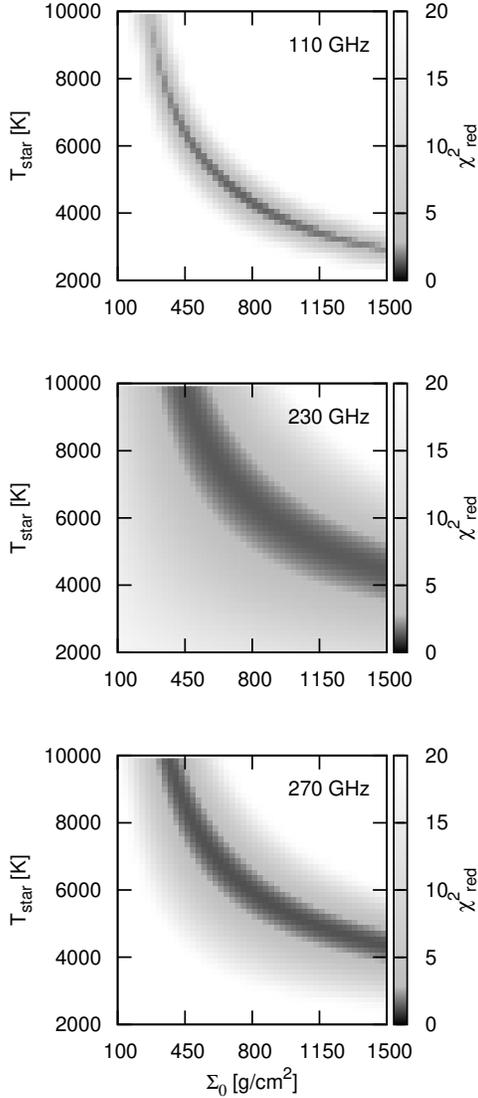}
  \captionstyle{normal}
  \caption{Fit criterion $\chi^2_{\rm red}$
 as a function of the normalization
of the surface density $\Sigma_0^{\rm gas}$
 and the temperature $T_{\rm star}$ of the
central star.}
  \label{Fig_Chi2SigmaT}
  \end{figure}

\begin{figure}[!Ht]
 \setcaptionmargin{5mm}
 \onelinecaptionsfalse
 \includegraphics[angle=0,width=0.7\textwidth]{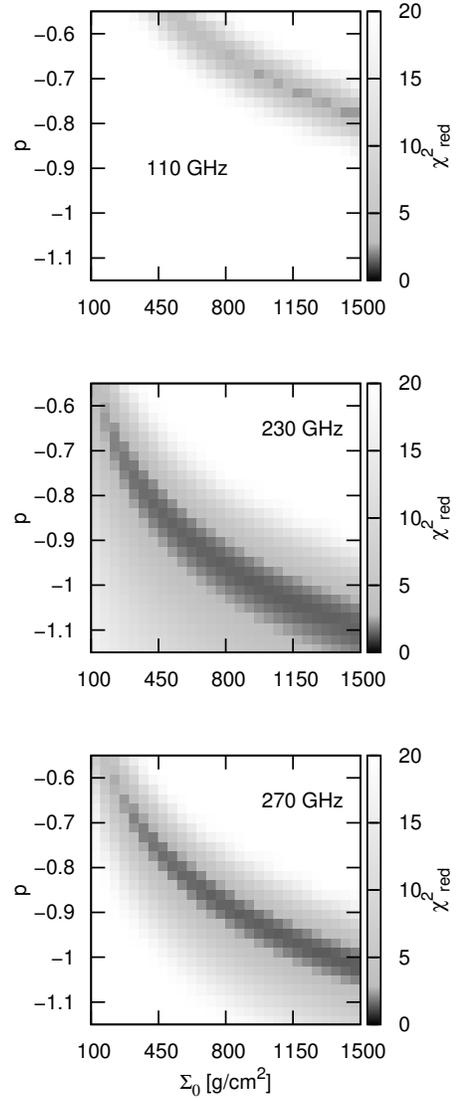}
 \captionstyle{normal}
 \caption{Fit criterion  $\chi^2_{\rm red}$
 as a function of the normalization
of the surface density $\Sigma_0^{\rm gas}$
and the index $p$ of the power-law
function for $\Sigma_{\rm T}$.}
 \label{Fig_Chi2Sigmap}
 \end{figure}

The fact that we observe the disk edge-on and
cannot see the direct radiation from the star hinders
determination of the spectral type of the central star.
The lack of information about the effective temperature
of the star means that the thermal structure of the disk must be reconstructed, and the characteristic
dust temperature required to determine the disk
mass~\cite{2005ApJ...631.1134A} must be estimated. There is a fundamental
(physical) constraint on the joint, simultaneous determination
of the mass and characteristic temperature
of the disk in the case of spatially unresolved
observations in the millimeter. We constructed maps
of $\chi^2_{\rm red}$ as functions of the structural parameters in
order to study the degeneracy of spatially resolved
observations. Figure~\ref{Fig_Chi2SigmaT} shows $\chi^2_{\rm red}$
as functions of
the normalization of the surface density $\Sigma_0^{\rm gas}$ and the
portion of the stellar radiation intercepted by the disk $f L_{\star}$. For ease of viewing, the latter quantity has been
transformed to the effective temperature of the star
using the standard formula
$f L_{\star}=f4\pi R_{\rm star}^2\cdot\sigma_BT_{\rm star}^4$ for $R_{\rm star}=R_{\odot}, f=0.1$. The values of parameters
from the dark-grey region with $\chi^2_{\rm red}\lesssim2$ (Figure~\ref{Fig_Chi2SigmaT}) describe
the observations equally well. This region of
parameters can be described by the law $\Sigma_0^{\rm gas}T^{\alpha}={\rm const}$,
where $\alpha=1.3\div1.7$. $\chi^2_{\rm red}$ behaves similarly
in the $\left(\Sigma_0^{\rm gas},p\right)$ parameter plane
(Figure~\ref{Fig_Chi2Sigmap}), with the
steeper decrease in the surface density toward the
periphery corresponding to greater values of $\Sigma_0^{\rm gas}$.

\subsection{Origin of the Differences of 110 GHz Map}
 \begin{figure*}
 \setcaptionmargin{5mm}
 \onelinecaptionsfalse
 \includegraphics[angle=270,width=0.8\textwidth]{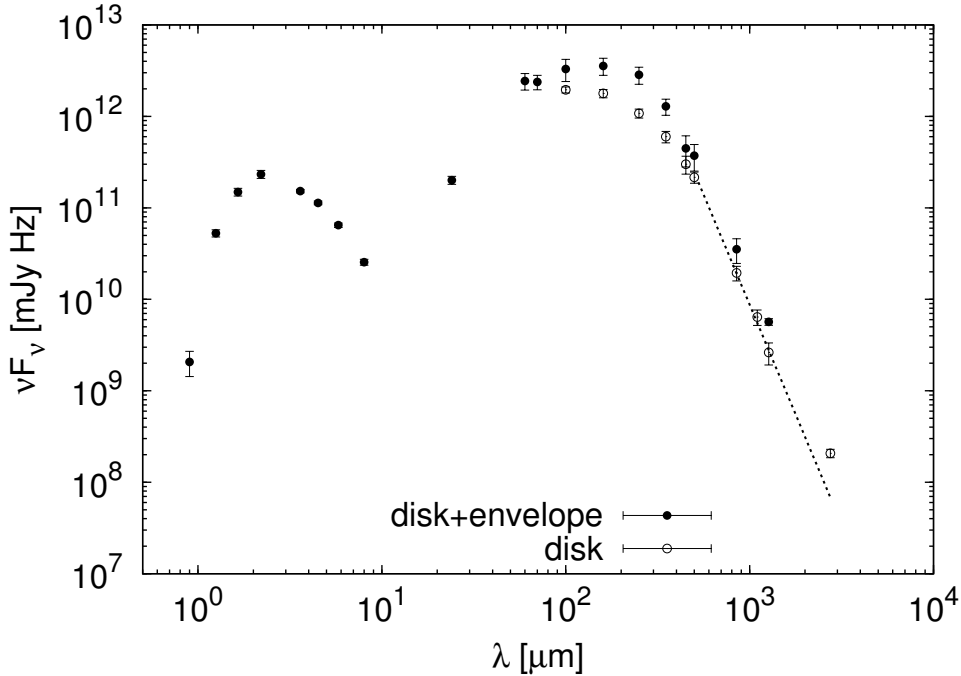}
 \captionstyle{normal}
 \caption{SED of CB 26. The open circles show the data from which the envelope's contribution has been subtracted. For ease of
viewing, the straight line shows the slope of the Rayleigh--Jeans part of the spectrum, derived using the last four points, except
for 110 GHz.}
 \label{Fig_SED}
 \end{figure*}
All the available data indicate that the 110 GHz
map (corresponding to longest of the wavelengths considered) differs from the remaining two. First, the
residuals display correlated features within the disk
image (Figure~\ref{Fig_MapOfTheMaps}), suggesting a systematic difference
between the observational and theoretical data at this
frequency, or that the derived $\chi^2_{\rm red}$ represents a local
minimum. Second, the disk parameters derived using
this map differ appreciably from those obtained using
the 230 and 270 GHz maps (Table~\ref{Tab_BestFits}).
Third, if the
point corresponding to 110 GHz is plotted on the
SED (Figure~\ref{Fig_SED}), it deviates appreciably from the straight
line in the Rayleigh--Jeans part of the spectrum. To
plot Figure~\ref{Fig_SED}
we used the fluxes from \cite{2009A&A...505.1167S} together
with data from the {\it Herschel}. The deviation
of a linear SED and the differences in the disk parameters at 110 GHz and at the other frequencies
can be explained in the following way:
\begin{itemize}
 \item there exists an additional radiative mechanism
that is not taken into account in the model, which is
not important at high frequencies, but becomes appreciable
at 110 GHz (e.g., free--free radiation arising
in the accretion region or bipolar outflow);
 \item an envelope around the disk could also contribute
to the radiation, since the 110 GHz beam
is larger than the 230 and 270 GHz beams (we are
planning to use CO line observations to estimate the
envelope's contribution);
 \item it is difficult to establish an accurate dust model
(chemical composition and extinction efficiency coefficients).
\end{itemize}

These (or other) effects could result in the overall
difference between the 110 GHz map and the maps
at 230 and 270 GHz. It seems reasonable to use
the results based on the latter two maps (rather than
all three) as a final disk model, until we determine
the nature of the operating additional mechanism and
take it into account.

  \section{CONCLUSIONS}

The appearance of modern radio interferometers
with high sensitivities and good angular resolutions
(SMA, Plateau de Bure, CARMA, ALMA) has made
it possible to obtain spatially resolved images of protoplanetary
disks. The reconstruction of the physical structures of the disks from these observations is
a challenging task that requires an integrated approach,
that is simultaneously self-consistent physically
and correct mathematically, aimed at modeling
and fitting the observations. One of the main goals of
this paper was to develop such an approach.

The key characteristic of the protoplanetary disk
model we used to reconstruct the disk parameters
is a compromise between the completeness of the
description provided and the time required to compute
the physical structure of the disk. The use of a
two-frequency approximation and the corresponding
mean opacities in the radiative transfer model, as
well as the iterative scheme employed to solve the
equation of hydrostatic equilibrium, yield a short
model computation time. Together with the rapid
computation of the theoretical maps, the adaptive
algorithm for convolution of the images with the
beam, and the efficient method used to minimize  $\chi^{2}_{\rm red}$,
this made it possible to use the developed code to
search for best-fit model parameters of the disk. 

We applied this code to determine the physical
structure of the protoplanetary disk in CB 26 using
interferometric images in the millimeter. Observational
data at 110, 230, and 270 GHz from the SMA,
IRAM Plateau de Bure, and OVRO interferometers
were used to search for best-fit models. Best-fit
model parameters were obtained for each of the frequencies,
for all the frequencies jointly, and for 230
and 270 GHz jointly. All three observational maps suggest the existence of a central, dust-free region
in the disk, approximately 35~AU in radius. This corresponds
well to the value of 45 AU obtained in~\cite{2009A&A...505.1167S} at 230~GHz. However, our simulated maps yield
a disk inclination of $78^{\circ}$,
which is lower than the
value of $85^{\circ}$ used in \cite{2009A&A...494..147L}. This should lead to an
appreciable decrease in the bipolar outflow velocity
for the model~\cite{2009A&A...494..147L}, and to a lower rate of angular momentum
transfer from the disk. We cannot
determine whether or not the dust particles in the
disk can grow appreciably in size (see also~\cite{2009A&A...505.1167S}); we
can only establish an upper bound on the dust size ($a_{\rm max}<0.02$~cm) using our dust model, comprising
a mixture of silicate and graphite particles with a
power-law size distribution.

We came across the following difficulties when
simulating CB 26. The 110 GHz map stands out from
the others due to the change in the SED slope, the
presence of symmetrical features in the residual map
within the disk image, and differences in the physical
parameters implied by the best-fit model. These may
indicate that (a) the model does not include free--free
radiation, although this may be comparable with the
thermal radiation of the disk at longer wavelengths;
(b) the contribution from a disk envelope has been neglected;
(c) the dust model is not sufficiently realistic.
All these factors must be considered in more detail.
Our analysis also indicates a wide region of degeneracy
between the thermal and density characteristics
of the disk. In our opinion, this is a fundamental problem
hindering our ability to unambiguously establish
the disk parameters. This problem can be avoided in
objects, in which the radiation of the central star is
not screened by the circumstellar disk, and the stellar
parameters can be determined independently. It is
also obvious that an increase in angular resolution
is necessary for this degeneracy to be removed. We
plan to study in future whether or not the region of
parameter degeneracy can be diminished using maps
from other wavebands.

  \begin{acknowledgments}
This work was supported by the Russian Foundation
for Basic Research (projects 10-02-00612, 12-02-33044, and 12-02-31248), the Federal Targeted
Program ''Scientific and Science Education Staff of
Innovational Russia'' for 2009--2013
(no. 14.B37.21.0251), and the Program of Support
for Leading Scientific Schools of the Russian Federation
(grant NSh-3602.2012.2). The authors thank
D.Z.Wiebe, M.S. Khramtsova, and S.Wolf for helpful
discussions. V.V. Akimkin thanks A.V. Brechalov
and A.A. Fedotov for valuable comments.
\end{acknowledgments}
  \bibliography{paper.bib}
\begin{flushright}
 {\it Translated by N. Lipunova}
\end{flushright}

\end{document}